# Predicting Atmospheric Re-entries Using Segmented M/A Calibration of Public TLE Data

**Joseph Remis,** Groupe Européen d'Observation Stellaire (GEOS), Saint-Avold, France
ORCID: 0000-0002-2438-5959
Email: joseph.remis@gmail.com
*Corresponding author*

## Abstract

We present a reproducible method for predicting atmospheric re-entries using only publicly available Two-Line Elements (TLEs). The approach is based on a segmented calibration of the effective mass-to-area ratio (M/A), adjusted independently over intervals of at least 12 hours to suppress short-period catalogue noise. Longer segments provide a stable dynamical signal from which the effective drag behaviour can be retrieved. Applied to recent natural re-entries of rocket bodies, debris, and non-manoeuvring satellites, the method consistently identifies the final orbital revolution and reproduces published Tracking and Impact Prediction (TIP) times with an accuracy of approximately ±1 hour in the terminal phase. The limiting factor is the intrinsic noise and cadence of the TLE catalogue rather than atmospheric or solar-flux modelling. Within these constraints, segmented M/A calibration offers a transparent, lightweight, and operationally robust alternative for re-entry analysis when only public data are available.

## 1. Introduction

Predicting the final hours of orbital decay is challenging because publicly available TLEs contain short-period noise that obscures the true dynamical evolution of low-altitude objects. Long-arc drag fitting accumulates this noise and often diverges, whereas short-interval anchoring to successive catalogue observations stabilises the reconstruction.

The physical state of a re-entering object, density environment, attitude, tumbling, thermal conditions, and ballistic coefficient, can vary on minute timescales, making detailed ab-initio modelling quickly obsolete. An empirical approach anchored directly to observed orbital evolution therefore provides a practical alternative.

The method developed here calibrates the effective mass-to-area ratio (M/A) over successive short propagation arcs, each anchored to the next TLE. This prevents long-arc divergence and isolates catalogue noise as the main uncertainty source. In the final hours before re-entry, the secular drop in semi-major axis becomes large enough to dominate residual noise, yielding hour-scale accuracy.

Moderate solar-flux variations have limited impact over such short intervals, and daily changes are usually small. The method is therefore largely insensitive to short-term solar activity.

This framework reconstructs the decay path directly from the semi-major-axis evolution encoded in successive TLEs, without requiring detailed knowledge of the object's geometry or aerodynamic state.

## 2. Data Used

The analysis uses only public data: successive Two-Line elements (TLEs) and preliminary and final Tracking and Impact Prediction (TIP) messages from Space-Track.

The achievable accuracy is constrained by the publication cadence, internal precision, and source diversity of the TLE catalogue. These structural limitations cannot be controlled by the analyst, and the method is designed to operate within them by extracting the dynamical signal directly from the semi-major-axis evolution.

All propagations were performed using NASA's GMAT with a fully specified force model: EGM96 gravity (21×21), NRLMSISE-00 atmospheric density, canonical solar-radiation pressure, constant drag and reflectivity coefficients, and IAU 2000/2006 Earth orientation parameters. The only adjusted parameter is the effective area $A$, which captures the net drag behaviour over each segment.

This empirical use of TLEs, treating them as a time-resolved sequence of orbital measurements, allows the decay to be reconstructed step by step despite the limited fidelity of the underlying data.

## 3. Physical Principles

Orbital decay in the terminal phase is governed primarily by atmospheric density, the effective mass-to-area ratio (M/A), and short-term variations in solar flux. At low altitudes, drag dominates all other perturbations.

Because the aerodynamic state of a re-entering object can change rapidly due to attitude, tumbling, thermal effects, and surface evolution, detailed physical modelling is impractical. These variations occur on timescales far shorter than they can be measured or predicted, and the segmented calibration absorbs their net effect into the effective area $A$, which acts as an operational proxy for drag behaviour. Only the ratio $M/A$ governs the drag acceleration, so the absolute mass need not be known.

Atmospheric density is computed using NRLMSISE-00. Short-term solar-flux variations produce modest density changes and do not significantly affect the calibration over the short intervals considered.

The effective mass-to-area ratio therefore captures the combined influence of geometry, attitude, thermal state, and unmodelled perturbations, without attempting to separate these contributions explicitly.

## 4. Methodology — Segmented M/A Calibration

The method reconstructs the terminal decay by calibrating the effective mass-to-area ratio $M/A$ over successive short propagation arcs anchored to the TLE catalogue. Each pair of TLEs defines an independent segment: the earlier TLE is propagated to the epoch of the later one, and the effective drag area $A$ is adjusted so that the simulated semi-major-axis decay matches the observed decay.

Let $a_1$and $a_2$be the semi-major axes extracted from two TLEs separated by $\Delta t$. The observed decay is

$$\Delta a_{\text{obs}} = a_2 - a_1.$$

Trial values of $A$ yield simulated decays $\Delta a_{\text{sim}}(A)$. Linear interpolation provides the optimal area:

$$A_{\text{opt}} = A_1 + (A_2 - A_1)\frac{\Delta a_{\text{obs}} - \Delta a_{\text{sim}}(A_1)}{\Delta a_{\text{sim}}(A_2) - \Delta a_{\text{sim}}(A_1)}.$$

A final propagation verifies that the residual is below 10 m. The effective mass-to-area ratio is then

$$(M/A)_{\text{eff}} = M/A_{\text{opt}}.$$

Successive values of $A$ form a time-resolved sequence tracing the object's drag behaviour throughout its decay. Long-arc divergence is avoided because each segment is anchored to a new catalogue measurement.

To suppress catalogue noise, a minimum separation of 12 hours between TLEs is adopted. In the terminal phase, when the secular drop in semi-major axis becomes large enough to dominate residual fluctuations, segments are accepted when the semi-major axis decreases by at least 10 km between epochs.

All propagations use a fixed GMAT force model; the only adjusted parameter is the effective area $A$.

### 4.1 Segment Stability Analysis

Table 1 summarises the effective areas obtained for object 68537 using successive and double-spaced TLE pairs.

Table 1

| EPOCH | A N-N+1 | A N-N+2 |
|---|---|---|
| 131.083951 | 7.5 | 7.5 |
| 131.516209 | 7.5 | 8.3 |
| 131.577941 | 8.6 | 8.1 |
| 131.824824 | 8.2 | 7.8 |
| 132.194998 | 7.9 | 7.9 |
| 132.68827 | 8.2 | 8.2 |
| 132.934772 | 8.5 | 8.0 |
| 133.181179 | 7.8 | 7.9 |
| 133.550604 | 8.3 | 8.2 |
| 133.858274 | 8.4 | 8.4 |
| 134.227224 | 8.6 | 8.3 |
| 134.473024 | 8.3 | 8.5 |
| 134.841453 | 9.1 | 8.2 |
| 135.086874 | 6.6 | 8.2 |
| 135.209525 | 8.6 | 8.9 |
| 135.760818 | 9.4 | 9.3 |
| 136.066507 | 9.2 | 8.8 |
| 136.554107 | 7.0 | 8.8 |

Successive-epoch values span a wider range (6.6–9.4 $m^2$), whereas double-spaced estimates cluster around 8–9 $m^2$. This contrast reflects the influence of short-period catalogue noise: closely spaced TLEs do not sufficiently decorrelate local fluctuations, while wider separations provide a more stable dynamical signal.

To ensure that each segment reflects genuine orbital evolution rather than catalogue noise, a minimum separation of 12 hours between TLEs is adopted. Over such intervals, the secular decay of the semi-major axis dominates residual fluctuations, allowing the effective drag behaviour to be retrieved reliably.

In the terminal phase, the situation changes. As the object approaches re-entry, the semi-major-axis drop becomes large enough to overwhelm short-period noise even when TLEs are closely spaced. In this regime, segments are accepted when the semi-major axis decreases by at least 10 km between epochs, ensuring that each calibration is anchored to a sufficiently strong dynamical signal.

This stability analysis supports the segmentation strategy: long enough intervals are required to suppress catalogue noise, while the final hours of decay naturally provide the necessary dynamical leverage.

## 5. Results

Only objects with final TIP messages of ±1 min accuracy and natural decay are included. For each case, the penultimate and final predictions derived from the last available TLEs are compared with the published TIP time.

Latest Predictions Compared with the Final TIPs

| NORAD | PredTime (UTC) | Prediction UTC | Final TIP UTC | TUD(h) | TIP - Pred (min) |
|---|---|---|---|---|---|
| 43802 | 01/06/2026 11:17 | 21/06/2026 14:17 | 21/06/2026 11:18 | 72 | -180 |
| 43802 | 20/06/2026 07:12 | 21/06/2026 10:29 | 21/06/2026 11:18 | 28 | 49 |
| 43802 | 21/06/2026 01:45 | 21/06/2026 09:39 | 21/06/2026 11:18 | 10 | 99 |

| NORAD | PredTime (UTC) | Prediction UTC | Final TIP UTC | TUD(h) | TIP-Pred (min) |
|---|---|---|---|---|---|
| 61440 | 13/06/2026 16:47 | 14/06/2026 16:58 | 14/06/2026 18:49 | 26 | 111 |
| 61440 | 14/06/2026 06:11 | 14/06/2026 16:53 | 14/06/2026 18:49 | 13 | 116 |
| 61440 | 14/06/2026 10:00 | 14/06/2026 17:48 | 14/06/2026 18:49 | 9 | 61 |
| | | | | | |
| 58465 | 09/06/2026 08:34 | 14/06/2026 16:19 | 14/06/2026 04:52 | 116 | -687 |
| 58465 | 11/06/2026 06:10 | 14/06/2026 09:50 | 14/06/2026 04:52 | 71 | -298 |
| 58465 | 13/06/2026 16:16 | 14/06/2026 06:56 | 14/06/2026 04:52 | 13 | -124 |
| | | | | | |
| 63227 | 09/06/2026 08:59 | 10/06/2026 06:07 | 10/06/2026 08:46 | 24 | 159 |
| 63227 | 09/06/2026 19:14 | 10/06/2026 06:24 | 10/06/2026 08:46 | 14 | 142 |
| 63227 | 10/06/2026 04:22 | 10/06/2026 07:11 | 10/06/2026 08:46 | 4 | 95 |
| | | | | | |
| 69179 | 06/06/2026 09:06 | 07/06/2026 12:54 | 07/06/2026 15:29 | 30 | 155 |
| 69179 | 07/06/2026 06:23 | 07/06/2026 12:48 | 07/06/2026 15:29 | 9 | 161 |
| 69179 | 07/06/2026 09:43 | 07/06/2026 15:02 | 07/06/2026 15:29 | 6 | 27 |
| | | | | | |
| 68363 | 04/06/2026 09:31 | 06/06/2026 13:51 | 06/06/2026 12:15 | 51 | -96 |
| 68363 | 05/06/2026 07:25 | 06/06/2026 10:57 | 06/06/2026 12:15 | 29 | 78 |
| 68363 | 06/06/2026 01:21 | 06/06/2026 12:54 | 06/06/2026 12:15 | 11 | -39 |
| | | | | | |
| 69093 | 28/05/2026 10:17 | 01/06/2026 15:07 | 01/06/2026 14:02 | 101 | -65 |
| 69093 | 31/05/2026 09:25 | 01/06/2026 10:56 | 01/06/2026 14:02 | 29 | 186 |
| 69093 | 01/06/2026 03:06 | 01/06/2026 12:36 | 01/06/2026 14:02 | 11 | 86 |
| | | | | | |
| 69050 | 17/05/2026 09:55 | 21/05/2026 14:30 | 22/05/2026 06:45 | 121 | -465 |
| 69050 | 20/05/2026 17:32 | 22/05/2026 04:10 | 22/05/2026 06:45 | 37 | 155 |
| 69050 | 21/05/2026 12:21 | 22/05/2026 04:34 | 22/05/2026 06:45 | 18 | 131 |
| | | | | | |
| 69094 | 17/05/2026 14:22 | 21/05/2026 13:38 | 18/05/2026 13:36 | 23 | -2 |
| 69094 | 18/05/2026 05:09 | 21/05/2026 11:50 | 18/05/2026 13:36 | 8 | 106 |
| | | | | | |
| 68537 | 14/05/2026 18:14 | 16/05/2026 21:36 | 16/05/2026 19:25 | 49 | -131 |
| 68537 | 15/05/2026 19:00 | 16/05/2026 21:50 | 16/05/2026 19:25 | 24 | -145 |
| 68537 | 16/05/2026 15:53 | 16/05/2026 18:23 | 16/05/2026 19:25 | 4 | 62 |
| | | | | | |
| 68838 | 26/04/2026 16:13 | 28/04/2026 21:52 | 27/04/2026 22:41 | 30 | -1391 |
| 68838 | 27/04/2026 14:40 | 27/04/2026 22:17 | 27/04/2026 22:41 | 8 | 24 |
| | | | | | |
| 20947 | 12/04/2026 05:48 | 17/04/2026 15:55 | 17/04/2026 11:31 | 126 | -264 |
| 20947 | 15/04/2026 07:55 | 17/04/2026 15:54 | 17/04/2026 11:31 | 52 | -263 |
| 20947 | 16/04/2026 18:26 | 17/04/2026 12:44 | 17/04/2026 11:31 | 17 | -73 |
| | | | | | |
| 68413 | 10/04/2026 08:35 | 13/04/2026 05:14 | 13/04/2026 03:40 | 67 | -94 |
| 68413 | 12/04/2026 13:03 | 13/04/2026 01:14 | 13/04/2026 03:40 | 15 | 146 |
| 68413 | 12/04/2026 18:28 | 13/04/2026 02:42 | 13/04/2026 03:40 | 9 | 58 |

Across all objects, the final prediction consistently reduces the deviation relative to the final TIP, often by a factor of two to four compared with the penultimate estimate. Convergence occurs within the last 6–12 hours of orbital life, when the semi-major-axis decay becomes large enough to dominate catalogue noise.

The method reliably identifies the final complete orbit prior to atmospheric entry, sharply constraining the geographical band potentially affected by the event and allowing all regions outside the final ground track to be excluded with certainty.

The predictions generated by the method are disseminated through:

• X (@jremis), a profile used almost exclusively for this purpose;

• the ReentryWatch group (IAC Abu Dhabi);

• satflare.com

### 6. Independent Observational Confirmations

Several re-entry events reconstructed with the segmented calibration were later confirmed through independent observations, providing external validation of the predicted trajectories and decay times.

#### PROGRESS MS-34 SL-4 R/B (27 April 2026)

A luminous atmospheric entry was observed from the International Space Station during docking operations. The reported timing and viewing geometry match both the reconstructed trajectory and the final TIP message for the SL-4 stage.

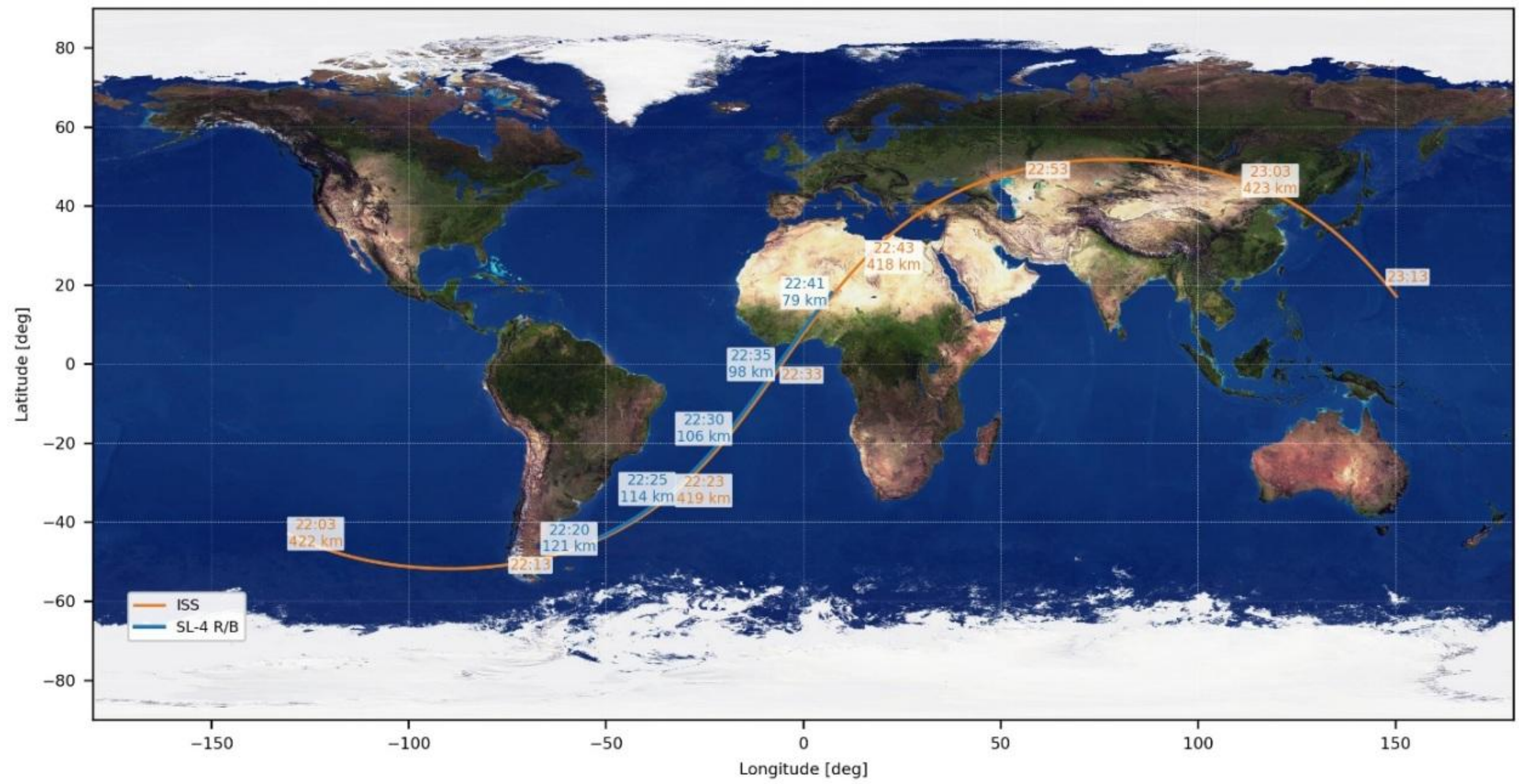

Figure 1. Ground tracks of the ISS and the SL-4 R/B stage on 27 April 2026.

#### ICEYE-X26 (31 May 2026)

Multiple witnesses in Sri Lanka reported a bright re-entry event. The observed timing aligns with the trajectory derived from the TLE sequence and is consistent with the final TIP.

### CLARITY-1 (10 June 2026)

Ninety-one reports collected by the American Meteor Society, including photographs and videos, correspond to the predicted ground track and final TIP for CLARITY-1. The event was also noted in the SpaceWeather daily bulletin.

These confirmations show that the segmented calibration reliably reconstructs the final orbital revolution and decay geometry using only public TLE data.

## 7. Starlink Constellation: Re-entry Behaviour

The Starlink constellation represents the largest population of active spacecraft in low Earth orbit and therefore accounts for a substantial fraction of recent atmospheric re-entries. This statistical weight makes it necessary to treat these satellites as a distinct class in re-entry studies.

Two broad categories are encountered:

- **Decommissioned satellites**, whose decay may or may not be accelerated depending on remaining propellant.
- **Prematurely non-operational satellites**, for which re-entry is typically accelerated unless the failure affects propulsion directly.

Three representative cases illustrate these behaviours.

Re-entry Prediction Summary for STARLINK-2171 (47754) *(compared to the final TIP)*

| Publication time | TLE epoch used | Predicted re-entry (UTC) | Uncertainty | Δ vs. Final TIP |
|---|---|---|---|---|
| 06 June 2026 – 15:57 UTC | 157.04458 | 08 June – 21:20 | ± 10 h | + 3 h 44 min |
| 08 June 2026 – 07:29 UTC | 159.01246 | 08 June – 15:48 | ± 2 h | – 1 h 48 min |
| 08 June 2026 – 09:54 UTC | 159.25703 | 08 June – 16:26 | ± 1.5 h | – 1 h 10 min |
| Final TIP | — | 08 June – 17:36 | ± 1 min | |

### Natural decay: STARLINK-2171 (47754)

This satellite exhibited no detectable thrusting. The segmented M/A calibration produced stable predictions that converged toward the final TIP as the object approached re-entry.

Re-entry Prediction Summary for STARLINK-1363 (45568) *(compared to the final TIP)*

| Publication time | TLE epoch used | Predicted re-entry (UTC) | Uncertainty | Δ vs. Final TIP |
|---|---|---|---|---|
| 01 June 2026 – 05:49 UTC | 151.94175 | 01 June – 11:02 | ± 2 h | – 9 h 16 min |
| 01 June 2026 – 15:44 UTC | 152.55264 | 01 June – 20:14 | ± 1 h | – 4 min |
| Final TIP | — | 01 June – 20:18 | ± 1 min | — |

### Accelerated decay with thrusting: STARLINK-1363 (45568)

Thrusting activity advanced the decay. A few hours before re-entry, the calibrated M/A values indicated that propulsion had ceased; the predicted decay time shifted later and matched the final TIP.

Re-entry Prediction Summary for STARLINK-34110 (63811) *(compared to the final TIP)*

| Publication time | TLE epoch used | Predicted re-entry (UTC) | Uncertainty | Δ vs. Final TIP |
|---|---|---|---|---|
| 02 June 2026 – 09:58 UTC | 152.75532 | 04 June – 04:27 | ± 9 h | + 1 day 13 h 44 min |
| Final TIP | — | 02 June – 14:43 | ± 1 min | |

**Rapid deorbit maneuver: STARLINK-34110 (63811)**

A sudden deorbit maneuver forced a rapid re-entry. The semi-major axis dropped from 6639 km to 6617 km in less than 48 hours, accompanied by a sharp increase in calibrated area (from 4–5 m² to more than 18 m²). Real-time prediction was impossible during the maneuver; only a post-event reconstruction could be performed.

Retrospective analysis of intermediate TLEs

| Date (UTC) | SMA (km) | A (m²) | M/A | Predicted Decay (UTC) |
|---|---|---|---|---|
| 30 May 15:26 | 6637.41 | 4.20 | 126.2 | 08 Jun 02:21 |
| 31 May 00:23 | 6635.63 | 4.79 | 110.6 | 07 Jun 00:53 |
| 31 May 15:19 | 6632.14 | 4.79 | 110.6 | 07 Jun 04:04 |
| 01 Jun 09:12 | 6627.33 | 18.33 | 28.9 | 02 Jun 20:27 |
| 01 Jun 18:08 | 6617.21 | 18.33 | 28.9 | 02 Jun 20:19 |
| Final TIP | | 23.40 | 22.6 | 02 Jun 14:43 |

The rapid drop in semi-major axis 6639 → 6617 km in < 48 h) indicates a highly energetic deorbit maneuver. The sudden increase to A ≈ 18.3 m² (M/A ≈ 28.9) marks the onset of thrusting. The predictions issued on 31 May, pointing to a 7 June re-entry, become irrelevant once the deorbit maneuver begins, and even the retrospective estimates remain imprecise as the thrusting appears to intensify (A rising from 18.33 m² to 23.4 m² when adjusted to the actual TIP). Although TLEs continued to be published normally, their anomalous behaviour could only be recognized afterwards, so no real-time prediction was possible during that interval.

## 8. Limitations

Catalogue noise, quantisation effects, and the simplified SGP4 dynamics impose an upper bound on the accuracy achievable with public TLE data. Short segments are particularly affected, as local fluctuations in the catalogue can dominate the observed decay and prevent stable drag estimation. Reliable calibration requires segments long enough for the secular evolution of the semi-major axis to exceed residual noise.

Controlled maneuvers or thrusting episodes temporarily invalidate short-term predictions. During such intervals, the calibrated drag parameters lose physical meaning because the observed semi-major-axis evolution reflects propulsion rather than atmospheric drag. Predictions become reliable again only once the orbit returns to a purely ballistic regime.

Even under favourable conditions, the final decay time can typically be recovered with an accuracy of about +1 hour. This limit reflects the intrinsic information content of the TLE catalogue rather than any deficiency in the modelling framework.

These constraints define the operational envelope of the segmented M/A calibration: robust reconstruction of natural decay, reliable identification of the final orbital revolution, and hour-scale accuracy in the terminal phase, provided that the object is not manoeuvring and that the TLEs reflect genuine dynamical evolution.

## 9. Discussion

The segmented M/A calibration provides a stable and reproducible framework for reconstructing terminal decay using only public TLE data. Anchoring each propagation arc to the next catalogue measurement prevents long-arc divergence and isolates catalogue noise as the dominant source of uncertainty.

Rapid aerodynamic variations, attitude changes, tumbling, thermal effects, and surface evolution, occur on timescales far shorter than they can be modelled. By absorbing these effects into the calibrated effective area, the method captures the net drag behaviour without requiring detailed physical assumptions. This empirical treatment is particularly important during the terminal phase, when the object's aerodynamic state evolves rapidly.

Across diverse re-entry scenarios, the method reliably identifies the final orbital revolution and matches published TIP times with hour-scale accuracy. The rapid convergence observed in the last hours of orbital life reflects the increasing dynamical signal contained in the semi-major-axis decay, which becomes large enough to dominate residual catalogue noise.

Overall, the performance of the method is limited not by atmospheric modelling or solar-flux variability, but by the intrinsic information content of the TLE catalogue. Within these constraints, segmented M/A calibration offers a lightweight, transparent, and operationally robust alternative to long-arc drag fitting when only public tracking data are available.

## 10. Conclusions

The segmented M/A calibration extracts the effective drag behaviour directly from the TLE catalogue and provides a practical and reproducible method for predicting atmospheric re-entries using public data alone. By enforcing segment lengths sufficient to suppress catalogue noise, the approach consistently recovers the final orbital revolution and achieves hour-scale accuracy in the final phase of decay.

Its minimal parameterisation, transparency, and reproducibility make it well suited for event identification, rapid decay reconstruction, and independent verification of publicly reported re-entries. Within the intrinsic limits of the TLE catalogue, the method offers a lightweight and operationally robust alternative to long-arc drag fitting when only public tracking data are available.

#### Declaration of competing interest

The author declares that he has no known competing financial interests or personal relationships that could have appeared to influence the work reported in this paper.

### Data Availability

All data used in this study are publicly available from Space-Track.org.

### CRediT authorship contribution statement

Joseph Remis: Conceptualization, Methodology, Software, Validation, Formal analysis, Investigation, Data curation, Visualization, Writing – original draft, Writing – review & editing.

### Acknowledgements

I am deeply grateful to Ennio Poretti for his thoughtful comments and expert guidance, to Juan Fabregat for his meticulous readings and invaluable suggestions, and extend special thanks to Alberto Buzzoni for his helpful comments on the first draft of this manuscript.